\documentclass[twocolumn]{jpsj2} 
\title{
Phase Diagram of the Triangular $t$-$J$ Model with Multiple Spin Exchange in the Doped-Mott Region
}

\author{Yuki \textsc{Fuseya}
\thanks{E-mail address: fuseya@hosi.phys.s.u-tokyo.ac.jp} 
and
Masao \textsc{Ogata}
}

\inst{
Department of Physics, University of Tokyo, 7-3-1 Hongo, Bunkyo-ku, Tokyo 113-0033
}

\newcommand{\bi}[1]{\ensuremath{\boldsymbol{#1}}}
\newcommand{\tjk}{$t$-$J$-$K$ }
\newcommand{\im}{{\rm i}}

\newcommand{\bq}{\boldsymbol{q}}

\usepackage{color}

\abst{We study the triangular $t$-$J$ model with multiple spin exchange interactions using exact diagonalization of small clusters, as a model of monolayer liquid $^3$He.
	A phase where the energy spectrum suggest the behaviors of spin-charge separation is realized in the doped-Mott region due to the competition between ferromagnetic two-spin and antiferromagnetic four-spin exchange interactions.
	This behavior gives an interpretation of the double peaked heat capacity recently observed in monolayer liquid $^3$He adsorbed on a graphite.
}

\kword{monolayer liquid $^3$He, doped-Mott insulator, triangular lattice, t-J model, multiple spin exchange, exact diagonalization, spin-charge separation}

\begin{document}
\maketitle

	Doping a Mott insulator opens up a new vista of the material physics\cite{Imada,Lee,Ogata}, which is similar to the evolution of the semiconductor physics in the last century.
	%
	%
	While the undoped semiconductor does not have the degree of freedom of either spin and charge, there remains the spin degrees of freedom in the Mott insulator.
	When hole carriers are doped in the Mott insulator, we should handle both the spin- and charge-degrees of freedom separately.
	This challenge is, however, not so easy.
	The problem of itinerancy and localization, which is very fundamental in quantum physics, are hidden behind this issue.
	%
	%
	%
	%
	%
	%
	%
	It is repeatedly examined, mainly in the context of high-$T_{\rm c}$ cuprates, whether the quasiparticle with spin and charge is well defined or not, which is still controversial\cite{Imada,Maekawa,Lee,Ogata}.
	%
	%
	%
	Theoretically, an interesting spin-charge separation has been addressed based on the $t$-$J$ model in square lattice\cite{Putikka,Maekawa}.
	Experimentally, the pseudo (spin) gap state suggests the spin-charge separation, but the direct indication of spin-charge separation has not been observed yet in the photoemission spectroscopy or in the heat capacity\cite{Ogata,Maekawa}.
	The difficulty to measure the spin-charge separation in high-$T_{\rm c}$ cuprates would be due to inhomogeneities or the close energy scale of spin and charge.
	So far the doped Mott system has been discussed mainly in the high-$T_{\rm c}$ cuprates.
	If we can approach the problem from a different standpoint with another system, we can obtain the essence of the doped-Mott systems more definitely.

	The second layer of $^3$He adsorbed on a graphite surface is another doped-Mott insulator in a purely two-dimensional system\cite{Greywall,Casey,Matsumoto05,Matsumoto}.
	In this system, $^3$He atoms localize at a certain density, the so-called 4/7 phase\cite{Greywall,Casey,Matsumoto,Matsumoto05,Ishida,Masutomi}.
	%
	%
	(We use the particle density, $n$, as $n=1$ at the 4/7 phase.)
	The effective mass exhibits a divergence toward the density of the 4/7 phase, $n=1$, indicating the Mott transition\cite{Casey}.
	We can, therefore, recognize this $^3$He system for $n \lesssim 1$ as the doped-Mott system.
	The unique features of this system are:
	(1) we can control the particle density without introducing disorder\cite{Greywall,Casey,Matsumoto05,Matsumoto};
	(2) the particles feel the triangular lattice potential\cite{Greywall,Elser};
	(3) the base Mott insulator does not exhibit any spin order, the gapless spin-liquid\cite{Ishida,Masutomi}, suggesting the energy scales of spin and charge are well separated.

	Recently, a double-peaked structure in the specific heat, $C$, has been reported in the doped-Mott region of monolayer $^3$He\cite{Matsumoto}.
	For low densities, $n < 0.85$, $C$ exhibits a single broad-peak (or a shoulder) at around $T_{\rm high}= 50$-100 mK, and $C\propto T$ below $T_{\rm high}$, indicating the Fermi liquid nature.
	When $n$ increases, a peak at around $T_{\rm low}= $1-10 mK appears for $n > 0.85$ in addition to the reduced peak at around $T_{\rm high}$.
	Then, $C$ shows a double-peaked structure for $0.85 < n <1$, namely, in the doped-Mott region.
	It is expected\cite{Matsumoto} that the double-peak structure originates from the separation of the energy scale of spin and charge, as in one-dimension\cite{Usuki}.
	(Hereafter we will use the term ``charge" following the custom of electron system.)
	%
	%
	%
	There have been, however, no justification of this scenario and we have no theories which explain how and why such a ``spin-charge separation" is realized in this system.
	%
	%
	%
	%
	
	In this letter, we present a theoretical justification of the behavior of ``spin-charge separation" in the doped-Mott region of monolayer liquid $^3$He.
	Here, we use the word ``spin-charge separation" in a broad sense, namely, it includes a situation where the energy scale of spin and charge is separated leading to a double-peaked structure of $C$, which might be described in the frame of the Fermi liquid theory.
	First an effective model is introduced;
	the $t$-$J$ model with the multiple spin exchange (MSE) interaction.
	Then the ground states of this model are investigated by the exact diagonalization.
	%
	We found a phase where the energy spectrum suggest the behavior of ``spin-charge" separation generated by a competition between MSEs.
	%

	%
	It is of prime importance to keep in mind the hierarchy of the energy scales, when we consider physical problems.
	Although $^3$He has a strong hard core, there is a process in which a $^3$He atom is excited to the third layer.
	%
	%
	%
	This process corresponds to the effective on-site repulsion $U$ of the Hubbard model\cite{Watanabe}, which is estimated as $U\sim 10$ K from the chemical potential difference between the layers\cite{Whitlock}.
	Our interest is below 100 mK, where the whole structure of double-peaked $C$ is included.
	In this case, we can use the $t$-$J$ model as an effective Hamiltonian as discussed in high-$T_{\rm c}$ cuprates\cite{Ogata}.
	%
	%
	%
	The hopping energy $t$ is estimated from the peak of $C$ to be $t\sim T_{\rm high}\sim 50$-100 mK\cite{Matsumoto}.

	It is well known that the MSE is important for hard-core quantum systems\cite{Thouless,Roger83,MisLhui}.
	%
	%
	%
	For example, for the monolayer solid $^3$He, $J_3 \gg J_2 \gtrsim J_4$ was estimated by Bernu et al.\cite{Bernu}, where $J_n$ is $n$-spin exchange interaction.
	Since $J_3$ term can be transformed into the $J_2$ term as $J=J_2-2J_3$, the sign of the effective two-spin exchange, $J$, is negative near $n=1$\cite{MisLhui,Roger98}.
	%
	%
	The magnitude of $J$ is estimated to be $|J| \sim T_{\rm low} \sim $1-10 mK.
	%
	%
	%
	The high-temperature expansion study yields $|J|\sim 3$mK, and the four-spin exchange $K$ to be positive and $K/|J|\sim 0.2$\cite{Roger98}.
	Consequently, the relevant energy scales below 100 mK are $t$, $J$ and $K$ which satisfy $t > |J|>K$.
	%
	%
	%
	%
	We therefore introduce an effective Hamiltonian of the monolayer liquid $^3$He on the basis of the triangular $t$-$J$ Hamiltonian with the MSE (four-spin) as follows:
\begin{align}
	H&=-t\sum_{i, j, \sigma}
	P_{\rm G}
\left(
 \,c_{i\sigma}^\dagger c_{j \sigma}
+ {\rm H.c.}
\right) P_{\rm G}
\nonumber\\
&+J\sum_{i, j}
\left(
\bi{S}_i \cdot \bi{S}_j
-\frac{n_i n_j}{4}
\right)
+K \sum \left(
P_4 + P_4^{-1}
\right),
\label{tJK}
\end{align}
	where $\bi{S}_i$ is the spin operator at site $i$, and $P_4$ is the four-spin exchange operators.
	The projection operator $P_{\rm G}=\prod_i \left( 1-n_{i\uparrow}n_{i\downarrow}\right)$ excludes the double occupancy, which reflects the hard-core potential, so that this \tjk Hamiltonian naturally expresses the hard-core quantum systems contrary to the Hubbard model.
	Near $n=1$, the effective two-spin exchange is ferromagnetic as $-J>K>0$.
	%
	%
	%
	At $n=1$, the \tjk model is equivalent to the previous MSE model ($J$-$K$ model) for the monolayer $^3$He\cite{MisLhui,Misguich,LiMing,Kubo,Momoi}.
	(Note that our definition of $J$ is twice as large as theirs.)
	%
	%
	%
	
	It should be noted here that the vacancies in $^3$He are treated as holes on the triangular lattice in the present model.
	Since the translational symmetry at 4/7 phase is different from that of the substrate\cite{Greywall}, the translational symmetry breaking occurs at a certain density.
	In order to consider such possibilities, it will be necessary to take account of the interstitial sites in the triangular lattice\cite{Watanabe}.
	In this paper, we use a simplified model Hamiltonian (\ref{tJK}) in order to see whether this minimal model exhibits ``spin-charge separation".
	%
	%
	%
	%
	Furthermore, some experiments\cite{Matsumoto} and recent numerical simulation\cite{Takagi} suggest that in the lightly-doped region, the holes move as if it keeps the same translational symmetry as that of the 4/7 phase.
	Therefore, we think the analysis based on the triangular lattice will distill the essence of the lightly-doped region of monolayer $^3$He.
	%
	%
	%
	%
	%
	%
	%

 \begin{figure}[tb]
 \begin{center}
 \includegraphics[width=7cm]{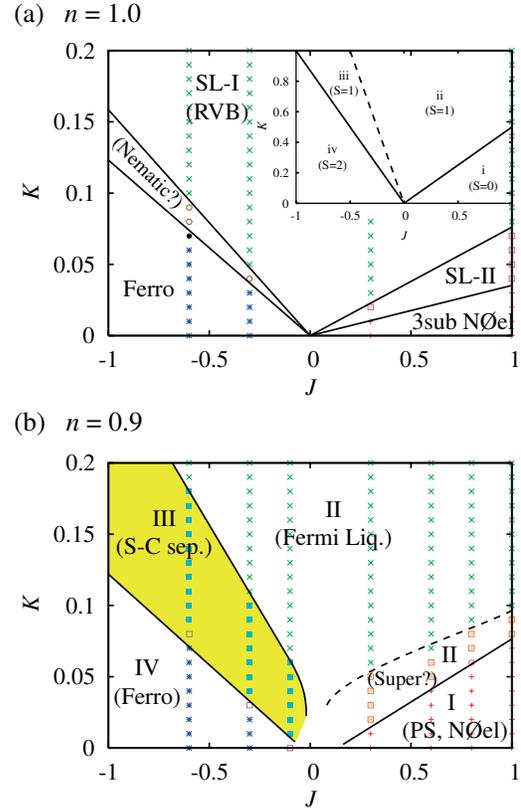}
 \end{center}
 \caption{
  (Color online).
 Phase diagram of the triangular \tjk model for (a) $n=1$ and (b) $n=0.9$ with $N_a =20$.
 The inset is the phase diagram of the four-spin plaquette.
 SL, S-C sep., and PS stand for spin liquid, possible ``spin-charge separation", and phase separation, respectively.
 \label{phase}}
 \end{figure}
	%
	%
	In order to obtain unbiased understandings of this model as a first step, we analyze the $t$-$J$-$K$ model using the exact diagonalization of small clusters.
	We have checked  12-, 18- and 20-site clusters with some cluster-shapes and periodic boundary conditions.
	We obtained a phase diagram, whose qualitative property is independent of the cluster size.
	Hereafter, we set $t=1$, and mainly show the results for $N_a=20$ site cluster.
	The phase boundaries are determined from the intersections of energy levels, and from the changes of spin- and charge-correlation functions, $S(\bi{q})$ and $N(\bi{q})$.
	%

	{\it At half-filling.}
	%
	First we briefly summarize the phase diagram for $n=1$.
	This filling has been studied up to 36-site for $K/|J|=1$\cite{Misguich} and $0.125$\cite{Momoi} with ferromagnetic (FM) $J$, and $K/J<0.125$ with antiferromagnetic (AF)  $J$\cite{LiMing}.
	Our results up to 20-site agree well with the previous works of 36-site not only qualitatively, but also quantitatively.
	Figure \ref{phase} (a) shows the phase diagram obtained for $n=1$.
	For $J>0$ with small $K$ $(<0.04J)$, the ground state is the three-sublattice N\'eel ordered state, which is verified by the relation\cite{Misguich,LiMing}
\begin{align}
	E\propto S(S+1)/N \chi ,
	\label{Neel}
\end{align}
    where $\chi$ is the susceptibility per site at zero field.
	In this N\'eel state, the spin correlation function
	%
$
	S(\bi{q})=N_a^{-1}\sum_{i, j} e^{\im \bi{q}\cdot\bi{r}_{ij} }
	\langle (n_{i\uparrow}-n_{i\downarrow})
	(n_{i\uparrow}-n_{i\downarrow})\rangle 
$
	%
	%
	shows peaks at the corners of the Brillouin zone of triangular lattice, the $K$-points, indicating the three sublattice structure.
	This three-sublattice N\'eel state is replaced by a spin liquid state for $0.04J < K < 0.08J$.
	This phase exhibits a large number of singlet excitations in the magnetic gap, which is called as the type-II spin liquid (SL-II) by Misguich and Lhuillier\cite{LiMing,MisLhui}.
	With larger $K$, the ground state changes into a different spin liquid state with a gap.
	This state corresponds to the type-I spin liquid of RVB (SL-I)\cite{Misguich,MisLhui}.
	%
	%
	The SL-I phase expands into FM region for $K> 0.15 |J|$.
	Finally between the FM and the SL-I, there is another state, which could correspond to the nematic state discussed by Momoi et al.\cite{Momoi}.
	%
	%

	{\it Doped-Mott region.}
	%
	Let us now go on to the main subject of this Letter, i.e., the doped-Mott region.
	The phase diagram for the doped-Mott region ($n=0.9$) is shown in Fig. \ref{phase} (b).
	In the region for $J>0$ with small $K$ ($<0.08J$), the phase-I, the ground state is phase separated, judging from the property of the compressibility $\kappa=(4N_a/N^2)/\{E(N+2)+E(N-2)-2E(N)\}$, where $E(N)$ is the ground state energy with $N$ fermions.
	%
	(As for $\kappa$, we calculated up to $N_a =18$.)
	This phase separation is similar to those obtained in one- \cite{Ogata91} and two-dimensional\cite{Yokoyama} $t$-$J$ model.
	%
	%
	%
	We also find that the relation of Eq. (\ref{Neel}) holds in the phase-I, indicating that the N\'eel order is formed in the phase-separated islands.
	%
	%
	Furthermore, $S(\bi{q})$ shows peaks at the $K$-points as the three-sublattice N\'eel phase of $n=1$.

	The phase separated N\'eel state is broken by $K$ and the ground state changes into the phase-II (and II').
	The relation of Eq. (\ref{Neel}) does not hold any more in the phase-II$^($'$^)$, so that the system turns into a homogeneous liquid.
	%
 \begin{figure}
 \begin{center}
 \includegraphics[width=7cm]{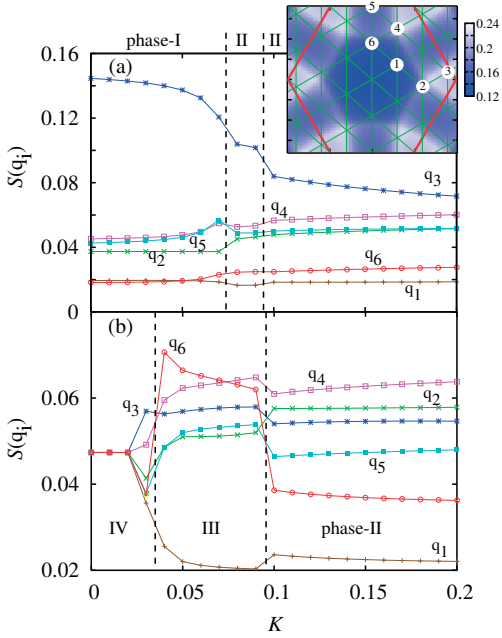}
 \end{center}
 \caption{
  (Color online).
 Spin correlation function for $\bi{q}_i$ as a function of $K$ for (a) $J=1.0$, and (b) $J=-0.3$.
 The inset shows the contour plot of $\chi_0 (\bi{q})$ and the position of $\bi{q}_i$.
 \label{Sq}}
 \end{figure}
	%
	The $K$-dependences of $S(\bi{q})$ for various $\bi{q}_i$ in the Brillouin zone are shown in Fig. \ref{Sq}.
	The positions of $\bi{q}_i$ are shown in the inset.
	For $J=1.0$ (Fig. \ref{Sq} (a)), the peak position, $\bq_3$, does not change from the phase-I to phase-II, but its magnitude becomes smaller.
	From this, we can see that the system possesses strong spin-fluctuations of the three-sublattice N\'eel order in the phase-II near the phase-I.
	As $K$ increases, however, the magnitude of $S(\bq_3)$ approaches to that of $S(\bq_4)$.
	This indicates that the system can be described as a normal Fermi liquid state, because the structure of $S(\bq)$ is similar to the non-interacting susceptibility 
$
	\chi_0 (\bi{q})=\sum_{\bi{k}} 
	\left(f_{\bi{k}+\bi{q}}-f_{\bi{k}}\right)
	/\left( \xi_{\bi{k}}-\xi_{\bi{k}+\bi{q}}\right) ,
$
	%
	where $f_{\bi{k}}$ is the Fermi distribution function and $\xi_{\bi{k}}$ is the fermion dispersion.
	(The contour plot of $\chi_0 (\bi{q})$ is shown in the inset of Fig. \ref{Sq}.)
	Namely, the spin fluctuation is reduced by increase of $K$.
	The energy spectrum also indicate the normal Fermi liquid property as shown later.
	%
	%
	The phase-II' would belong to the phase-II, since the property of $S(\bi{q})$ is almost the same.
	Note that there is a possibility that a superfluid is realized in the phase-II', since $\kappa$ is divergently large in this region.
	This situation is similar to the $t$-$J$ model\cite{Ogata91,Yokoyama}.
	If so, the mechanism of the superfluid can be due to the spin-fluctuation of three-sublattice N\'eel state.
	%
	
	A new phase (phase-III) is realized between the phase-IV of FM and the phase-II of normal Fermi liquid\cite{phase-V}.
	%
	%
	As is shown in Fig. \ref{Sq} (b) for $J=-0.3$, the structure of $S(\bi{q}_i)$ in the phase-III is definitely different from those of the phase-II and IV.
	%
	%
	The symmetry of the wave function of the phase-III, therefore, is different from the normal Fermi liquid in II and the FM in IV.
	This is contrary to the relation between the phase-I and II.
	%
	%
	%
	%

 \begin{figure}
 \begin{center}
 \includegraphics[width=7cm]{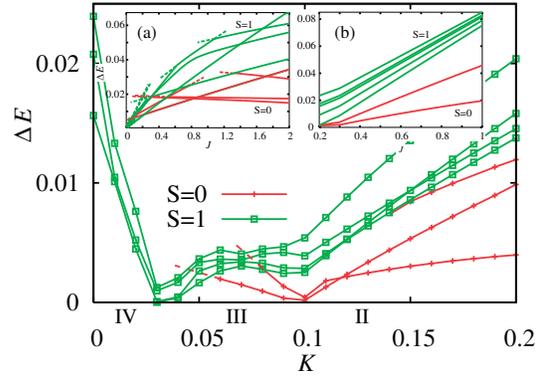}
 \end{center}
 \caption{
 (Color online).
 Excitation spectrum $\Delta E$ for $n=0.9$ with $J=-0.3$.
 The inset (a) is $\Delta E$ in a one-dimensional $t$-$J$ model for $n=0.875$ with $N_a=16$, as a reference of the Tomonaga-Luttinger liquid.
  The inset (b) is $\Delta E$ in a triangular $t$-$J$ model for $n=0.33$ with $N_a =12$, as a reference of the Fermi liquid.
 \label{excitation}}
 \end{figure}
	%
	The most specific feature of the phase-III is its excitation spectrum.
	The excitation energy $\Delta E$ of $S=0$ and $S=1$ are shown in Fig. \ref{excitation} for $J=-0.3$.
	In the phase-II ($K>0.1$), $\Delta E_{S=0}$ varies together with $\Delta E_{S=1}$.
	This is a characteristic property of the typical Fermi liquid (see the inset (b) of Fig. \ref{excitation}).
	In the phase-III ($0.04<K<0.1$), on the other hand, the energy spectrum behaves differently.
	The valley of $\Delta E_{S=0}$ locate at $K\sim 0.1$, while that of $\Delta_{S=1}$ at $K\sim 0.04$.
	%
	If the system is a normal Fermi liquid, the whole of $\Delta E$, including $\Delta E_{S=0}$, will decrease toward the FM state as the inset (b) of Fig. \ref{excitation}.
	However, the charge excitation $\Delta E_{S=0}$ in the phase-III rather increases toward the FM state, whereas the spin excitation $\Delta E_{S=1}$ keeps decreasing.
	In this sense, the charge sector is separated from the spin sector.
	%
	This behavior strikingly resembles that of the Tomonaga-Luttinger liquid, which definitely exhibits the spin-charge separation\cite{Maekawa,Ogata91}.
	(See the inset (a) of Fig. \ref{excitation}.)
	We speculate that this behavior strongly supports the ``spin-charge separation" in the phase-III.
	%
	From the structures of $S(\bi{q})$ and $\Delta E$, we conclude the phase-III is a new phase, where spin excitations are lowered and charge ones are raised.
	In such a situation, the heat capacity will exhibit the double-peaked structure, where the lower peak corresponds to the energy scale of the spin excitation and the higher one to that of the charge excitation\cite{Usuki}.
	Note that the state in the phase III is different from the Tomonaga-Luttinger liquid in the respect that the clear sign of spin-charge separation in the momentum space cannot be seen.
	Actually, we find no significant peaks both in $S(\bi{q})$ and $N(\bi{q})$ (not shown).
	%
	For the classification whether the phase-III is the non-Fermi liquid or not, we need further investigation.
	%

 \begin{figure}
 \begin{center}
 \includegraphics[width=8cm]{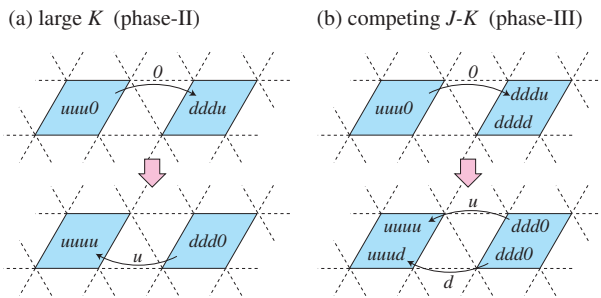}
 \end{center}
 \caption{
  (Color online).
 Illustration of the hole motion between the four site plaquette (the shaded region) for (a) the large $K$ region and (b) the competing $J$ and $K$ region.
 \label{intuitive}}
 \end{figure}
	%
	For obtaining the physical origin of the phase III, it is helpful to consider a superlattice of 4-site\cite{Misguich}.
	%
	%
	The eigenstates of the local four-spin plaquette are as follows:
	$E=-J+2K$ (for $S=0$), $J, 3J, J-2K$ (for $S=1$), $5J+2K$ (for $S=2$).
	Its phase diagram is shown in the inset of Fig. \ref{phase}.
	In the region of (i), the ground state is $S=0$. 
	In (ii), all $S=1$ states are lower than the others, in (iii) $S=2$ state locates between $S=1$ states, i.e., $E_{S=2}$ is almost degenerate with $E_{S=1}$, and in (iv) $S=2$ state is the ground state.
	The qualitative property is very similar to the phase diagram of $N_a=20$, which tells the plaquette approximation is effective.

	Let us first consider the large $K$ case.
	Within the plaquette, $K$ favors the $S=1$ state, i.e., the $uuud$ state.
	In the bulk limit of $n=1$, the $uuud$ plaquettes aligns AF with forming the RVB state as $\left| uuud, dddu - dddu, uuud\right.\rangle$.
	The hole is doped as $uuu0$, not as $uud0$, since $K$ is no longer valid in the plaquette.
	When the hole moves to the neighbor plaquette, the previous plaquette becomes $uuuu$ state (Fig. \ref{intuitive} (a)).
	The hole leaves a trace of spin as $uuuu$, $dddd$, $uuuu$..., which has an energy loss causing the spinon-holon binding\cite{Ogata}.
	This will be the situation in the phase-II, where the normal Fermi liquid is realized by doping in the RVB state.
	Near the FM state, on the other hand, AF $K$ compete with FM $J$, so that the ground states are highly degenerate\cite{Kubo,Momoi}.
	In such a situation, the $uuuu$ is the alternative to $uuud$ in the plaquette.
	Even when the hole moves to the neighbor plaquette, the previous plaquette keeps the degenerate state of $uuud$ and $uuuu$ (Fig. \ref{intuitive} (b)), namely, the hole does not leave a trace of energy-loss configurations.
	This is the very situation of the ``spin-charge separation".
	The spin excitation comes from the excitation between $uuud$ and $uuuu$ state, while the charge one from the motion of holes.
	Therefore, the key to the ``spin-charge separation" is the degeneracy of $uuud$ and $uuuu$ state (the competing $J$ and $K$).
	In other words, doping a gapless spin-liquid produces the ``spin-charge separated" liquid.
	In this Letter, the \tjk model was introduced as an effective model of the monolayer liquid $^3$He.
	A new phase (phase-III) was found near the FM phase in the doped-Mott region.
	The competition between the FM $J$ and AF $K$ generates the phase of possible ``spin-charge separation".
	The triangular \tjk model therefore naturally unravel the mysterious double-peaked heat capacity in monolayer liquid $^3$He.
	Our results suggest that doping a gapless spin-liquid yields a spin-charge separated state.
	These results open up a new possibility of the doped-Mott systems.
	%
	%
	%
	%
	%
	%
	%
	%
	%
	%
	%

	%
	The authors greatly appreciate Hiroshi Fukuyama for stimulating discussions and useful advice.
	Thanks are also due to K. Kubo and T. Momoi for useful discussions.
	%
	%
	This work was financially supported by Grant-in-Aid for Scientific Research on Priority Areas No. 17071002, and Next Generation Supercomputing Project, Nanoscience Program from MEXT.
	Y. F. is supported by JSPS Research Fellowships for Young Scientists.



\begin{thebibliography}{99}
\bibitem{phase-V}
	Note that there is a small region of another $S=0$ or partial FM between the phase-III and IV (indicated by open squares in Fig. \ref{phase} (b)), which seems to vanish in the thermodynamic limit.
\end{thebibliography}
\end{document}